\documentclass[letterpaper, 10 pt, conference]{ieeeconf}  

\IEEEoverridecommandlockouts                              
\usepackage[utf8]{inputenc}
\usepackage[T1]{fontenc}
\usepackage  {amsmath}
\usepackage{amssymb}
\usepackage{verbatim}
\usepackage[letterpaper,vmargin=3cm]{geometry}
\usepackage{tikz}
\usepackage {float}
\usetikzlibrary{matrix,calc,shapes}
\tikzset{
  treenode/.style = {shape=rectangle, rounded corners,
                     draw, anchor=center,
                     text width=5em, align=center,
                     top color=white, bottom color=blue!20,
                     inner sep=1ex},
  decision/.style = {treenode, diamond, inner sep=0pt},
  root/.style     = {treenode, font=\Large, bottom color=red!30},
  env/.style      = {treenode, font=\ttfamily\normalsize},
  finish/.style   = {root, bottom color=green!40},
  dummy/.style    = {circle,draw}
}

\title{\LARGE \bf
A Dynamic Quantum Clustering Approach to Brain Tumor Segmentation 
}

\author{Jacksson Sánchez$^{1}$ and Miguel Martín-Landrove$^{2}$
\thanks{This work was supported by the Universidad Nacional Pedro Henríquez Ureña, Santo Domingo, República Dominicana, Universidad Central de Venezuela, Caracas, Venezuela and the Physics and Mathematics in Biomedicine Consortium, P$\&$MBioC}
\thanks{$^{1}$J. Sánchez is with Faculty of Science and Technology, Physics Department,
        Universidad Nacional Pedro Henríquez Ureña, Santo Domingo, República Dominicana.
        {\tt\small jsanchez@unphu.edu.do}}%
\thanks{$^{2}$M. Martín-Landrove is with the Center for Medical Visualization,  National Institute for Bioengineering (INABIO), Universidad Central de Venezuela, Caracas, Venezuela and Centro de Diagnóstico Docente Las Mercedes,Caracas, Venezuela.
        {\tt\small mglmrtn@gmail.com}}%
}

\begin{document}

\maketitle
\thispagestyle{empty}
\pagestyle{empty}

\begin{abstract}

Data clustering has been widely used in data analysis and classification. In the present work, a method based on dynamic quantum clustering is proposed for the segmentation and analysis of brain tumor MRI. The results open the possibility of applications to multi modality medical imaging.

\end{abstract}

\section{INTRODUCTION}

Clustering of data is, in general, an ill-defined problem. Nonetheless it is a very important one in many scientific
and technological fields of study. Given a set of data-points one looks for possible structures by sorting out which
points are close to each other and, therefore, in some sense belong together. This is a preliminary stage taken before
investigating what properties are common to these subsets of the data. Dynamic quantum clustering \cite{c1}\cite{c2} has been proposed for the analysis of Big data \cite{c10} and machine learning \cite{c11} in an effort to understand relationships among data subsets. The sensitivity of the method to detect the appropriate number of classes within the  data has been used on previous applications to medical images \cite{c3}\cite{c4}. In the present work a full use of the dynamic quantum clustering algorithm is proposed for medical image segmentation.

\section{METHODS}

\subsection{Image Selection}

Images for high grade gliomas were extracted from different on line collections in The Cancer Imaging Archive \cite{c5}; The
Cancer Genome Atlas Low Grade Glioma (TCGA-LGG) data collection \cite{c6}, the Repository of Molecular Brain Neoplasia
Data (REMBRANDT) \cite{c7} for astrocytomas and oligodendrogliomas of grades 2 and 3, and The Cancer Genome Atlas
Glioblastoma Multiforme [TCGA-GBM] collection \cite{c8} for glioblastoma multiforme. For benign brain tumors, local image
datasets were used. Among these collections, T1-weighted images, either contrast enhanced or not, were selected and
further reviewed. Tumor lesions selected for image processing were clearly identified as such and separated from
anatomical structures.

\subsection{Dynamic Quantum Clustering Algorithm and image space}
\subsubsection{General}
The Dynamic Quantum Clustering Algorithm, DQCA, assumes that data are described by a set of points, each one defined with some uncertainty, $\sigma$ and the distribution for all points in data space is given by a Parzen estimator, $\varphi$, which satisfies the time independent Schrödinger equation for its ground state \cite{c1}\cite{c2},

\begin{equation}
 -\frac{1}{2\sigma^{2}}\nabla ^{2}\varphi+ V(\overrightarrow{X})\varphi = E\varphi = 0    
\end{equation}
\\
which leads to the evaluation of an energy potential,

\begin{equation}
 V(\overrightarrow{X})= \frac{1}{2\sigma^{2}} \frac {\nabla ^{2}\varphi} {\varphi}     
\end{equation}
\\
Applied to digital levels, the Parzen estimator is given by the convolution of the image histogram, $h$, with the gaussian kernel,

\begin{equation}
     \varphi(\overrightarrow{X})={\sum_{l=1}^{N} h(\overrightarrow{X_l}) e^{-\frac{(\overrightarrow{X}-\overrightarrow{X_l})^{2}}{2\sigma^{2}} }}
\end{equation}
\\
where the sum runs over $N$ digital levels, $\overrightarrow{X_l}$, each one with dimension $n$, i.e., the number of registered images that contribute to the information in an image element.The potential energy can be generally calculated as,

\begin{equation} 
\scriptstyle V(\overrightarrow{X})= \scriptstyle \frac{1}{2\sigma^{4}}\left[\frac{1}{\sigma^{2}} \frac{\sum_{i=1}^{N} h(\overrightarrow{X_l}) (\overrightarrow{X}-\overrightarrow{X_i})^{2}e^{-\frac{(\overrightarrow{X}-\overrightarrow{X_i})^{2}}{2\sigma^{2}} }}{\sum_{i=1}^{N} e^{-\frac{(\overrightarrow{X}-\overrightarrow{X_i})^{2}}{2\sigma^{2}} }}  - n  \right]    
\end{equation}
\\
\subsubsection{Dynamics}

Using the Ehrenfest theorem, it is possible to obtain the equation of motion for each digital level defined as,

\begin{equation}
    \frac{d^2\overrightarrow{X_k}(t)}{dt^2}=-\frac{\nabla V(\overrightarrow{X_k}(t))}{\sigma^2}
\end{equation}
\\
In order to induce clustering of the data to the potential energy minima, a dissipation term must be added so a second order Langevin equation is obtained, \cite{c9}

\begin{equation}
    \sigma^2\frac{d^2\overrightarrow{X_k}}{dt^2}=-\nabla V(\overrightarrow{X_k})-\gamma\frac{d\overrightarrow{X_k}}{dt}
\end{equation}
\\

The clustering process is performed with a suitable selection of parameters such as $\sigma^2$, which is an equivalent to a 'mass', dissipation, $\gamma$ and time interval. These parameters are estimated according to the magnitude of the digital levels and as a consequence depend on image data. An example of clustering of digital levels is shown in Figure 1.
\begin{figure}[H]
    \centering
    \includegraphics[width=7cm]{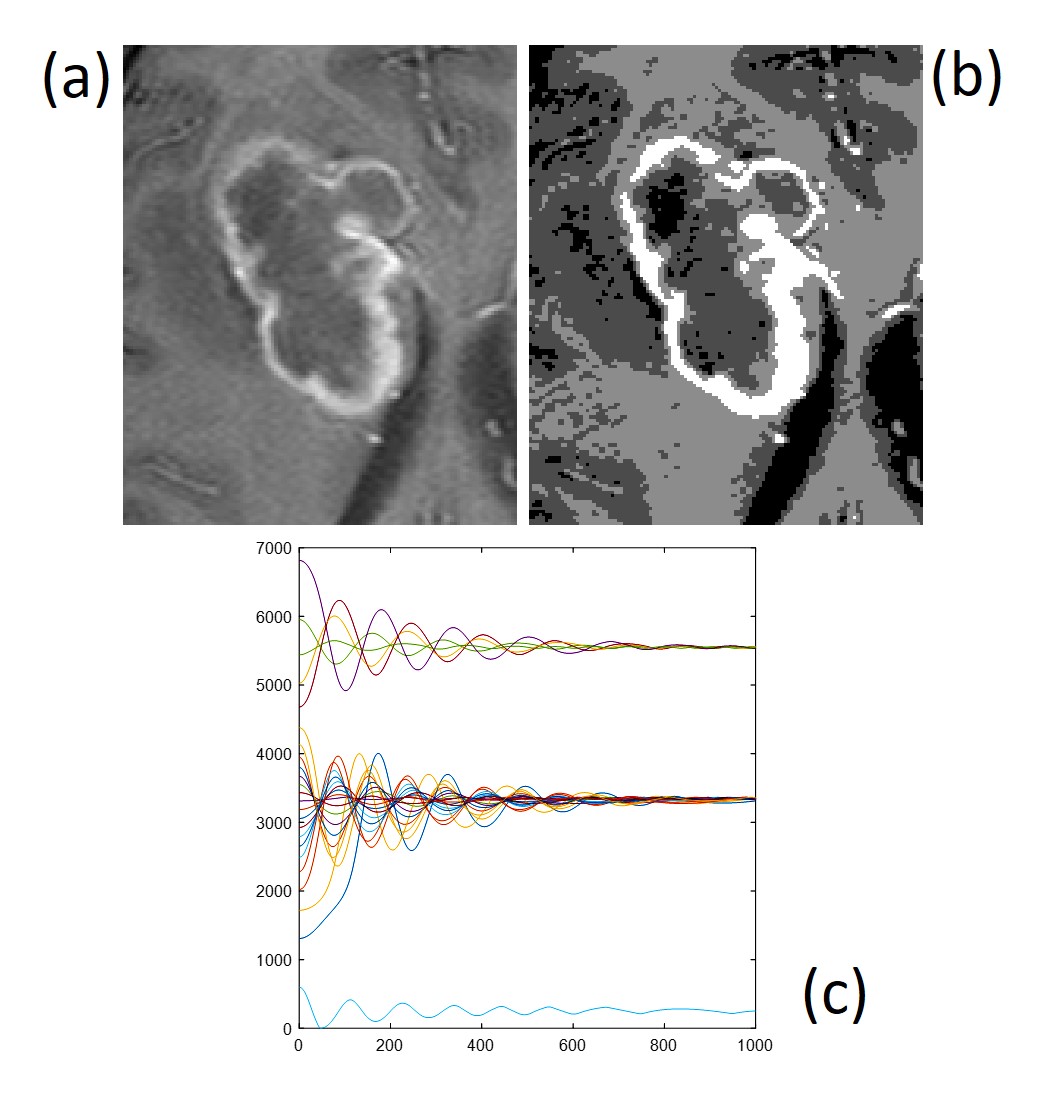}
    \caption{Clustering of digital levels. (a) Original image, (b) clustered image, (c) Time evolution of digital levels }
    \label{fig:f1}
\end{figure}
\subsection{Image Segmentation}

The general schematics is shown in Figure 2. Image segmentation is performed in 2 phases, Phase I, which performs a gross segmentation of the original image and eliminates unwanted anatomical structures and image artifacts, and Phase II, which is used to produce a refined segmentation. 

\begin{figure} [H]
    \centering
    \includegraphics[width=7cm]{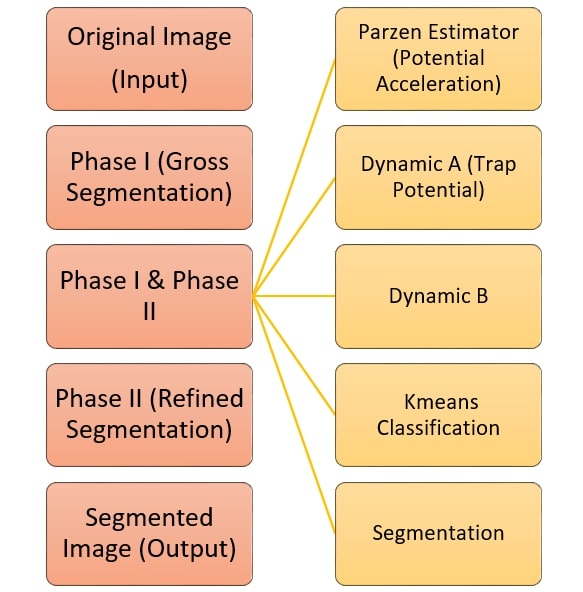}
    \caption{Schematic process of the proposed Dynamic Quantum Clustering Algorithm.Phases I and II have a common structure that is represented on the right hand side}
    \label{fig:f3}
\end{figure}

The structure of each phase is the following,
\begin{itemize}
    \item Parzen estimator, $\varphi(\overrightarrow{X})$
    \item Potential, $V(\overrightarrow{X})$
    \item Acceleration, $-\frac{\nabla V(\overrightarrow{X_k}(t))}{\sigma^2}$
    \item Dynamic A: First application of the dynamics upon the original image using the potential and acceleration previously calculated. At the end of this step, a Parzen estimator is evaluated using the clustered image histogram, allowing for the calculation of a 'trap' potential, $V_{trap}(\overrightarrow{X})$, which defines the number of classes to be used in a further classification algorithm, such as Kmeans.
    \item Dynamic B: Second application of the dynamics upon the original image using the 'trap' potential and the corresponding acceleration.
    \item Classification of the original and clustered images by Kmeans algorithm. The number of classes is determined by the number of minima in the 'trap' potential. Classified images are inspected to select those classes that are of interest. This step requires a qualified user. This step generates a segmentation mask that summarizes both the selected classes of the original and clustered images.
    \item Segmentation of the original image. The segmented image is further inspected to eliminate non interesting anatomical structures or possible image artifacts, as in the previous step, it requires a qualified user. In the case of Phase I, this step generates a 3-dimensional ROI to be used as input image for Phase II, which reduces digital level space for a more precise analysis and segmentation.
\end{itemize}

\section{RESULTS AND DISCUSSION}
As it was previously mentioned, the dynamic quantum clustering algorithm requires an appropriate set of parameters to be used in the dynamics, in particular, the selection of $\sigma^2$, $\gamma$ and the time interval depends on the properties of the image space, and a wrong selection of these parameters could produce unfinished clustering and classification of the images. The number of classes depends on the minima of the potential \cite{c2} and it is modified by the selection of $\sigma^2$. This matter can be seen in the example shown in Figure 3, as $\sigma^2$ is increased there is a reduction of the number of minima in the 'trap' potential with noticeable modifications in the segmented image. 

\begin{figure}[H]
    \centering
    \includegraphics[width=6.5cm]{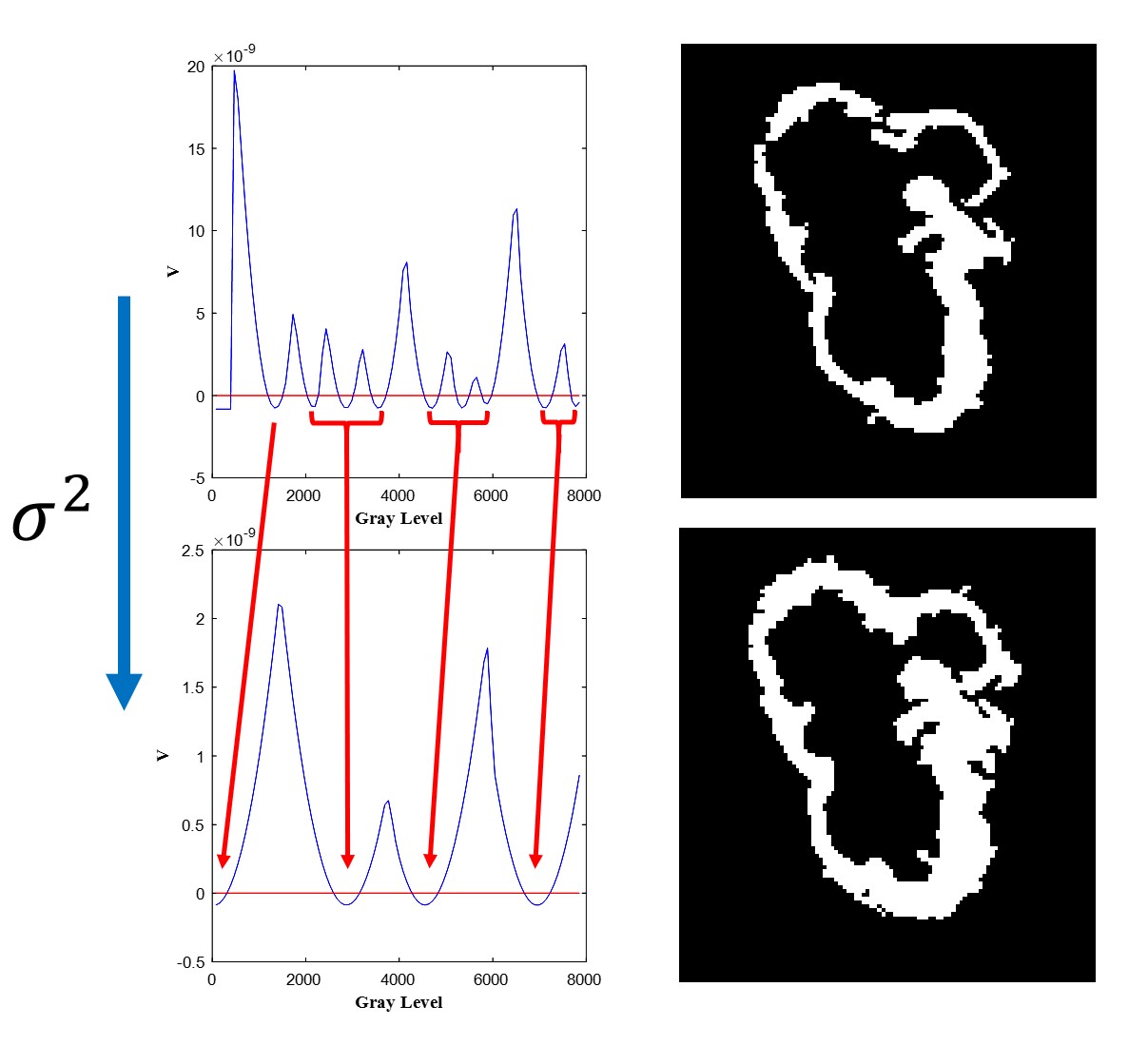}
    \caption{Dependence of the segmentation procedure on the value of the 'mass' of the particle, $\sigma^2$. Arrows indicate how the potential energy minima, in the 'trap' potential, collapse as $\sigma^2$ is increased from top to bottom by a 3-fold factor. On the right, the corresponding segmentations}
    \label{fig:f2}
\end{figure}

The selection of $\sigma^2$ is therefore a compromise, if it is too small, it becomes unpractical since the user must select more classes in order to segment the image, expending more time and expertise on it. Otherwise, if it is too large, the Parzen estimator becomes a unimodal function which completely destroys all the information in the image histogram. 
In the present work, $\sigma^2$ was selected proportional to the bin width of the image histogram and so it depends on the digital levels range. The average behavior of the number of classes with $\sigma$ is shown in Figure 4. Once $\sigma^2$ is selected, the Parzen estimator, potential and acceleration are determined straightforwardly. Time interval and $\gamma$ are then selected to provide an smooth pace of the dynamics as shown in the example of Figure 1.

\begin{figure}[H]
    \centering
    \includegraphics[width=6cm]{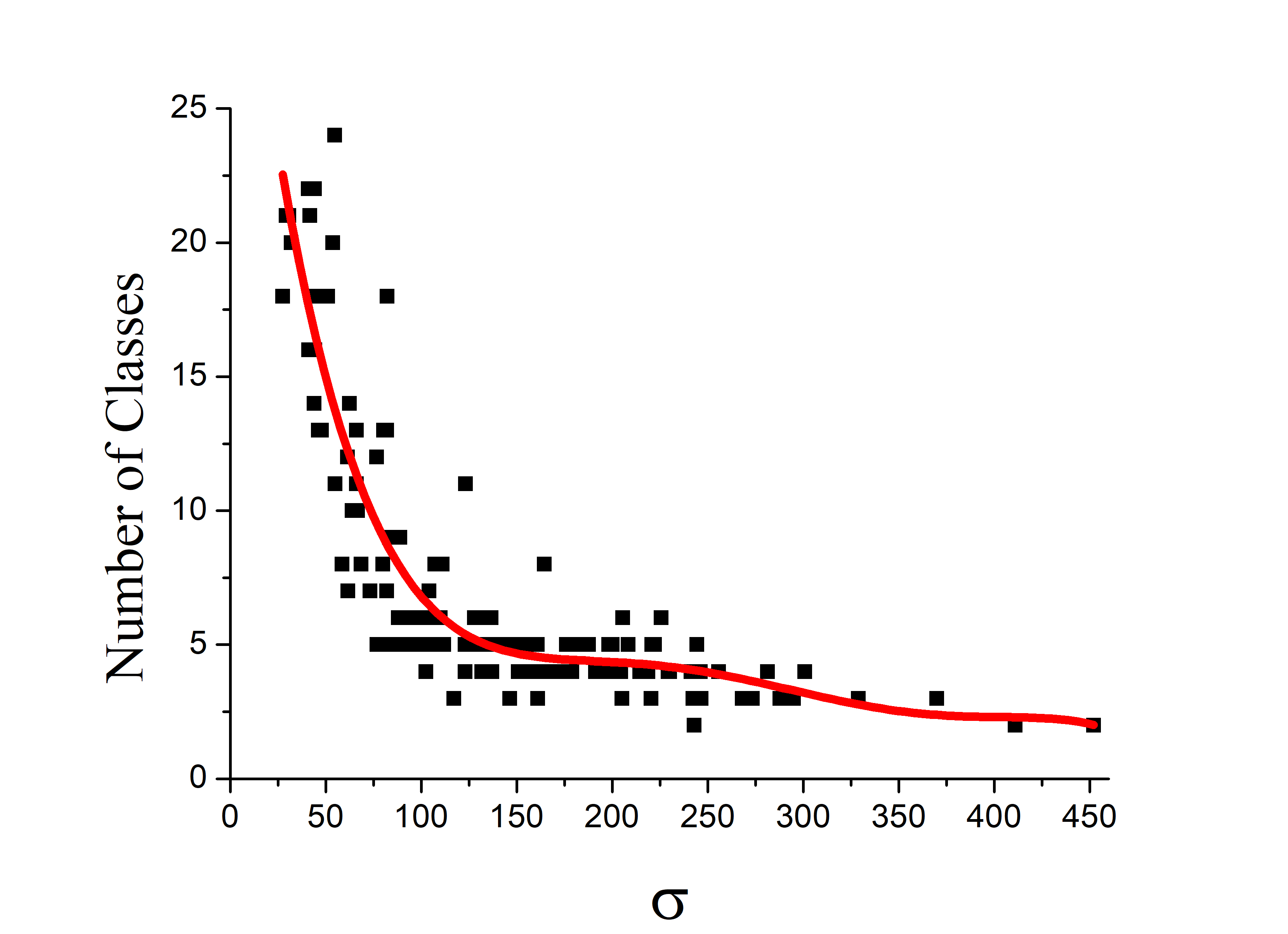}
    \caption{Dependence of the number of classes with $\sigma$. Red line represents the trend of the data}
    \label{fig:f5}
\end{figure}

The values of the 'trap' potential minima given by the dynamic quantum clustering algorithm exhibit a close correspondence to the class centroids obtained by a Kmeans algorithm with the same number of classes as can be seen in Figure 5, which was generated for an intermediate value of $\sigma$. This result is a consequence of the effectiveness of the clustering algorithm in guiding the digital levels to any neartby minimum of the 'trap' potential.

\begin{figure}[H]
    \centering
    \includegraphics[width=6cm]{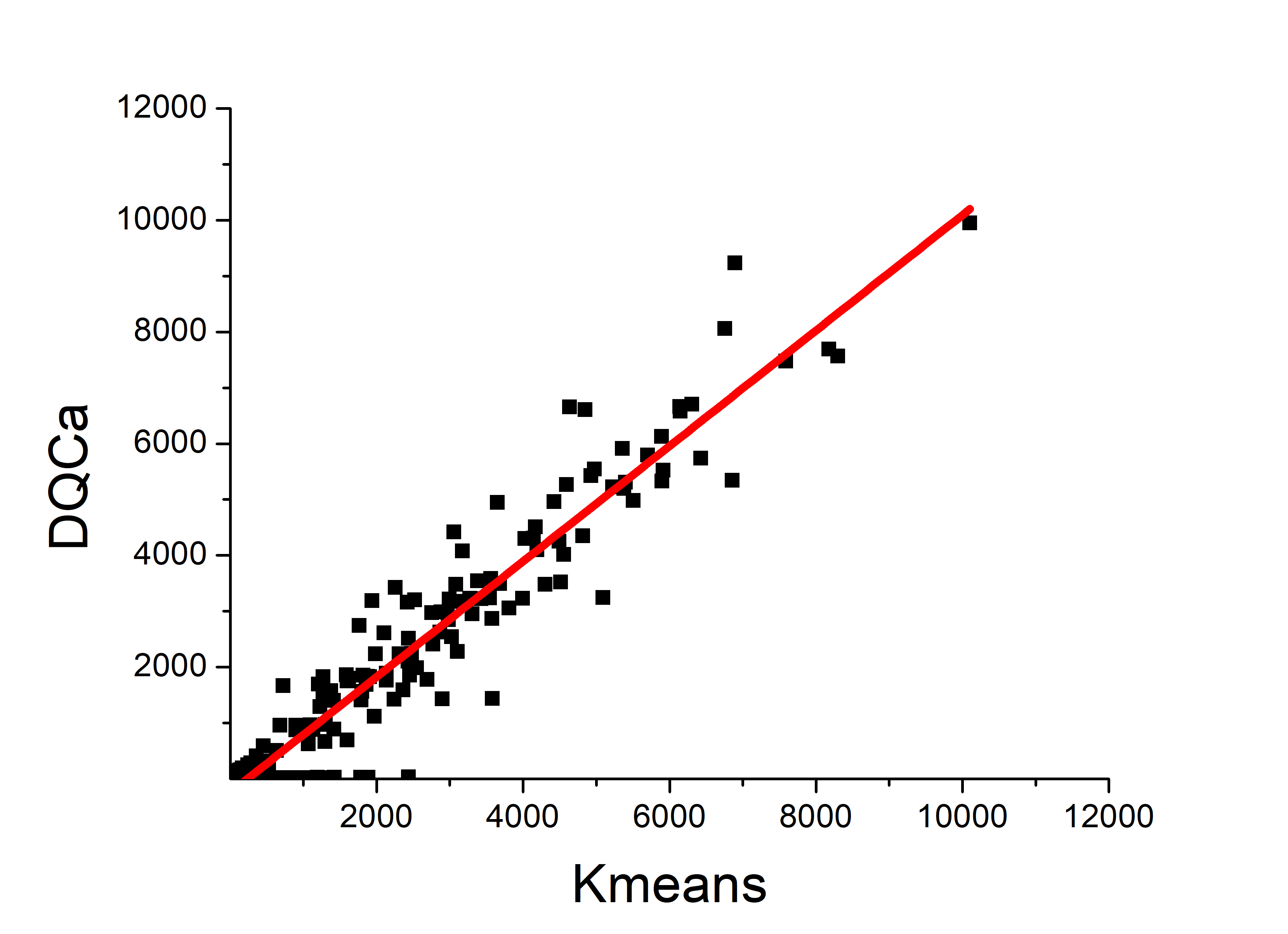}
    \caption{Correlation between the class centroids obtained by the Kmeans algorithm, and the 'trap' potential minima obtained by the Dynamic Quantum Clustring algorithm. The slope of the curve is $1.03 \pm 0.02$.}
    \label{fig:f6}
\end{figure}

When the two methods, Kmeans and DQC, are compared in the classification of the images, their similarity is quite good, as can be seen in Figure 6. 

\begin{figure}[H]
    \centering
    \includegraphics[width=5.5cm]{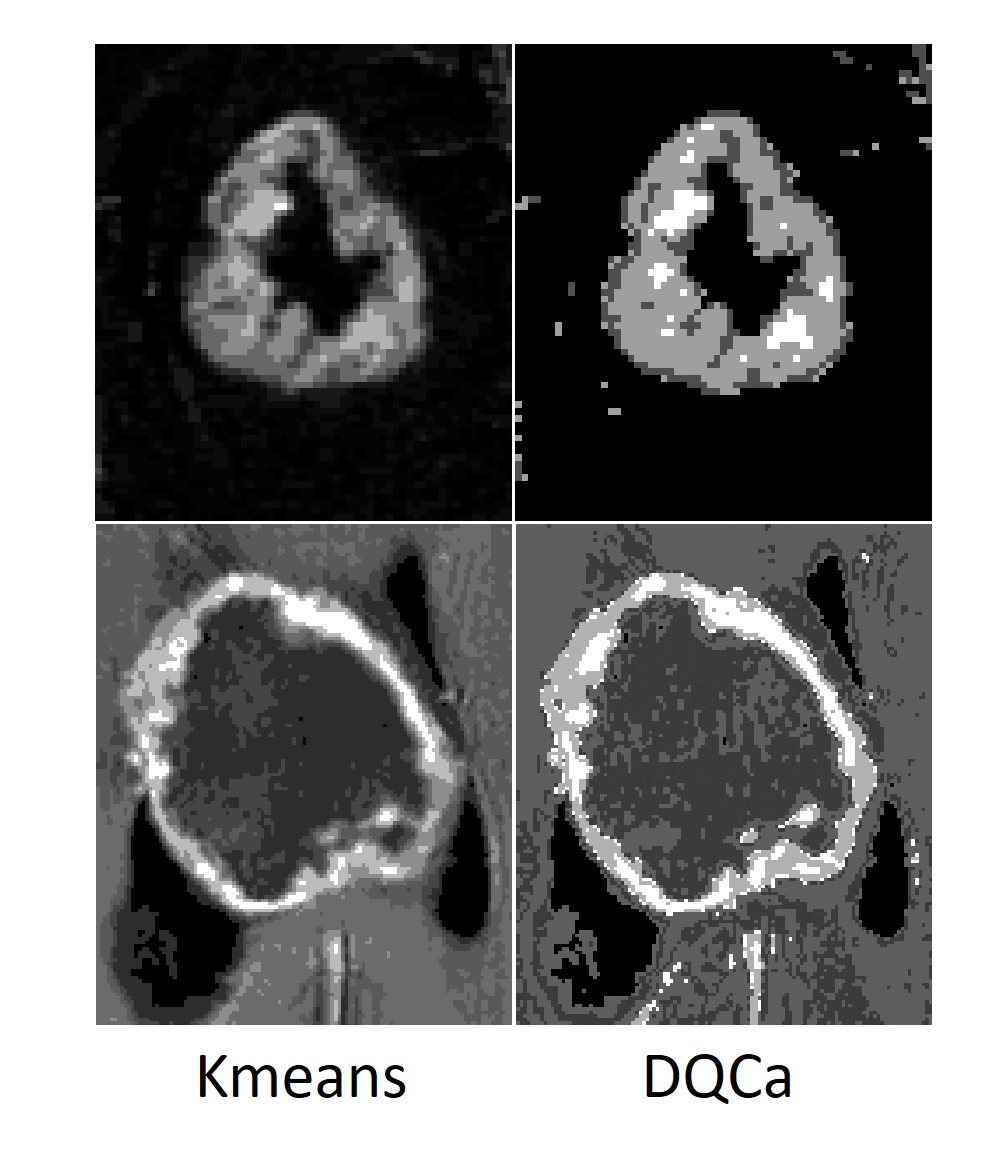}
    \caption{Examples of segmentation using the Kmeans algorithm, left column, and the Dynamic Quantum Clustering algorithm, right column}
    \label{fig:f7}
\end{figure}

This result suggest the possibility of using DQC alone for the segmentation of the image after its inspection by a qualified user, or establish some criteria for class selection. In any case, the elimination of unwanted anatomical structures or possible image artifacts still requires of a qualified user or equivalent. 

\begin{figure}[H]
    \centering
    \includegraphics[width=5.5cm]{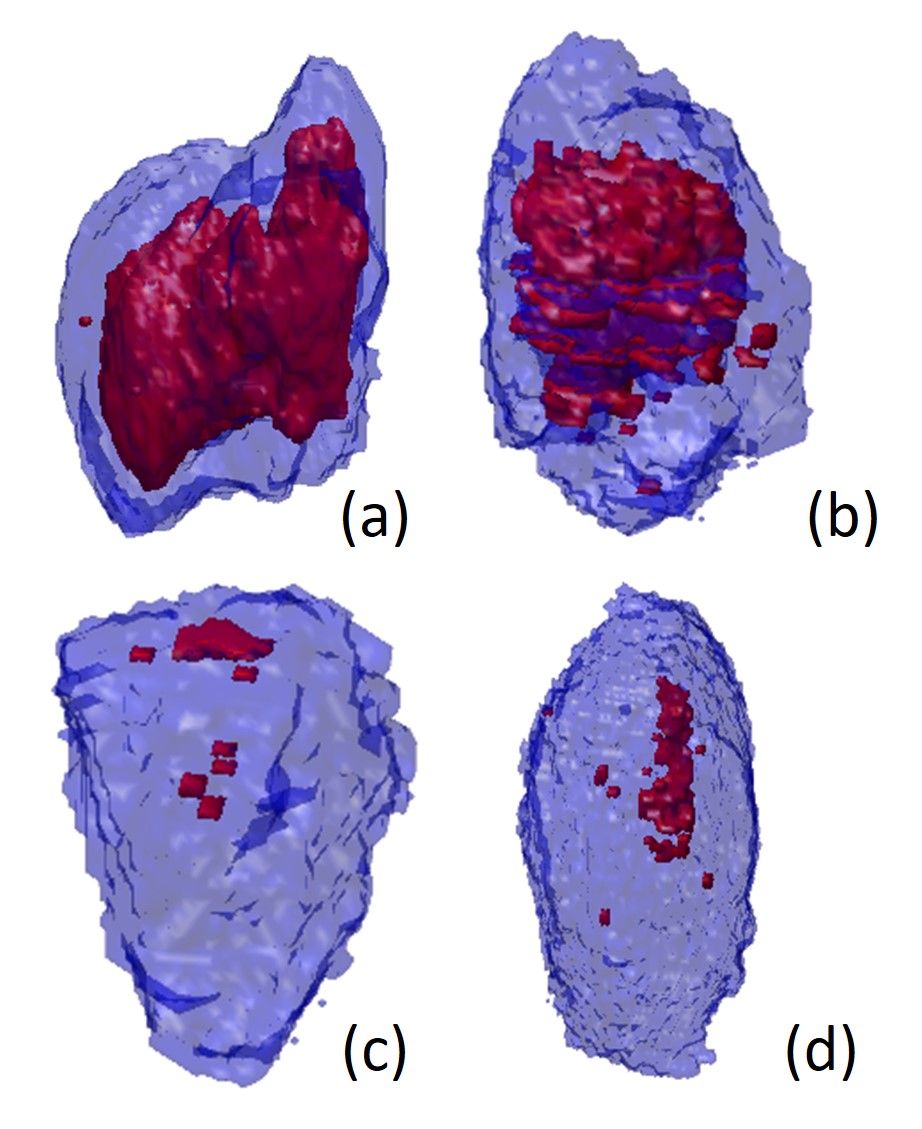}
    \caption{Examples of segmentation. (a) Glioblastoma multiforme,(b) High grade glioma, (c) Meningioma and (d) Acoustic Schwannoma. Blue color indicates enhanced contrast regions while red color indicates necrotic regions}
    \label{fig:f4}
\end{figure}

In Figure 7, it is shown an example of a detailed three dimensional reconstruction of tumoral lesions, which includes all the tumor types studied in this work.

\section{CONCLUSIONS}

In the present work, an effective method based on the dynamic quantum clustering algorithm was proposed for image segmentation of medical images. Although the research was performed on images of the same modality, the work can be easily extended to multimodality images. Further is undergone to fully automatize the process.

\addtolength{\textheight}{-12cm}

\section*{ACKNOWLEDGMENT}

This research was supported by Universidad Central de Venezuela, INABIO and Universidad Nacional Pedro Henríquez Ureña. We thank our colleagues from these institutions who provided insight and expertise that greatly assisted the research, although they may not agree with all of the interpretations/conclusions of this paper.

\end{document}